
\input vanilla.sty

%

%

\def\\{{\tt\char'134}}
\def\{{{\tt\char'173}}
\def\}{{\tt\char'175}}
\def\@{\char'100}
%
%
%
\def\boxit#1{%
              \vbox{%
                    \hrule%
                    \hbox{%
                          \vrule%
                          \kern3pt
                          \vbox{\kern3pt\hbox{#1}\kern3pt}%
                          \kern3pt%
                          \vrule%
                          }%
                     \hrule%
                    }%
             }
\newdimen\hautbox \newdimen\largbox
\newbox\entbox
\def\shadowbox#1{
                 \setbox\entbox=\boxit{#1}
                 \hautbox=\ht\entbox
                 \largbox=\wd\entbox
                 \advance \hautbox by -2pt
                 \vbox{\hbox{\boxit{#1}\vrule width2pt height\hautbox}
                 \nointerlineskip\moveright 2pt \hbox{\vrule height2pt
                 width\largbox}\nointerlineskip}}
\def\border#1{%
              \vbox{%
                    \hbox{%
                          \kern3pt
                          \vbox{\kern3pt\hbox{#1}\kern3pt}%
                          \kern3pt%
                          \vrule%
                          }%
                     \hrule%
                    }%
             }
%

%
%
\def\fonthdg{c}

\def\aujourdhui{\space\number\day\space\ifcase\month\or
janvier\or f\'evrier\or mars\or avril\or
mai\or juin\or juillet\or ao\^ut\or septembre\or
octobre\or novembre\or d\'ecembre\fi
\space\number\year}

\def\today{\space\ifcase\month\or
January\or February\or March\or April\or
May\or June\or July\or August\or September\or
October\or November\or December\fi
\space\number\day,\space\number\year}
\def\it#1{{\sl #1\/}}

%
%
\newcount\nosection
\newcount\nosubsect
\newcount\nosubsub
\nosection=0
\def\section#1#2{\advance \nosection by 1 \nosubsect=0 \nosubsub=0
               \removelastskip\bigskip\leftline{\moyen #2{\number
               \nosection\ #1}}
               \medskip}
\def\subsection#1#2{\advance \nosubsect by 1 \nosubsub=0
               \removelastskip\medskip
               \leftline{\bf\hskip\parindent #2{\number\nosection.\number
               \nosubsect\ #1}}
               \medskip}
\def\subsubsection#1#2{\advance \nosubsub by 1
                  \removelastskip\medskip
                  {\bf\hskip\parindent #2{\number\nosection.\number\nosubsect.
                  \number\nosubsub\ #1}}
                  \medskip}
%
%
%

%
=\fonthdg mr8
=\fonthdg mtt10
=\fonthdg mr6
=\fonthdg mmi8 \skewchar\eighti='177
=\fonthdg mmi6 \skewchar\sixi='177
=\fonthdg msy8 \skewchar\eightsy='60
=\fonthdg msy6 \skewchar\sixsy='60
=\fonthdg mbx8
=\fonthdg mbx6
=\fonthdg msl8
\font\eightit=\fonthdg mti8
=\fonthdg mcsc10


\def\Tenpoint{\normalbaselineskip=12pt            
 \def\rm{\fam0\tenrm}%
 \def\it{\fam\itfam\tenit}%
 \def\sl{\fam\slfam\tensl}%
 \def\bf{\fam\bffam\tenbf}%
\def\smc{\tenrm}
 \def\mit{\fam 1}%
 \def\cal{\fam 2}%
 \textfont0=\tenrm   \scriptfont0=\eightrm   \scriptscriptfont0=\sixrm
 \textfont1=\teni    \scriptfont1=\eighti    \scriptscriptfont1=\sixi
 \textfont2=\tensy   \scriptfont2=\eightsy   \scriptscriptfont2=\sixsy
 \textfont3=\tenex   \scriptfont3=\tenex     \scriptscriptfont3=\tenex
 \textfont\itfam=\tenit
 \textfont\slfam=\tensl
 \textfont\bffam=\tenbf \scriptfont\bffam=\eightbf
   \scriptscriptfont\bffam=\sixbf
\setbox\strutbox=\hbox{\vrule height 8.5pt depth 3.5pt width 0pt}%
\def\tt{\tentt}\normalbaselines\rm}

\def\Eightpoint{\normalbaselineskip=10pt
 \def\rm{\fam0\eightrm}%
 \def\it{\fam\itfam\eightit}%
 \def\sl{\fam\slfam\eightsl}%
 \def\bf{\fam\bffam\eightbf}%
 \def\mit{\fam 1}%
 \def\cal{\fam 2}%
 \textfont0=\eightrm   \scriptfont0=\sixrm   \scriptscriptfont0=\sixrm
 \textfont1=\eighti    \scriptfont1=\sixi    \scriptscriptfont1=\sixi
 \textfont2=\eightsy   \scriptfont2=\sixsy   \scriptscriptfont2=\sixsy
 \textfont3=\tenex   \scriptfont3=\tenex     \scriptscriptfont3=\tenex
 \textfont\itfam=\tenit
 \textfont\slfam=\eightsl
 \textfont\bffam=\eightbf \scriptfont\bffam=\eightbf
   \scriptscriptfont\bffam=\sixbf
\setbox\strutbox=\hbox{\vrule height 7pt depth 3pt width 0pt}%
\normalbaselines\rm}
=\fonthdg mr10 scaled \magstep1
=\fonthdg mtt10 scaled \magstep1
=\fonthdg mmi10 scaled \magstep1 \skewchar\twli='177
=\fonthdg msy10 scaled \magstep1 \skewchar\twlsy='60
=\fonthdg mbx10 scaled \magstep1
=\fonthdg msl10 scaled \magstep1
\font\twlit=\fonthdg mti10 scaled \magstep1

\def\twlpoint{\Twlpoint}
\def\Twlpoint{\normalbaselineskip=14.4pt            
 \def\rm{\fam0\twlrm}%
 \def\it{\fam\itfam\twlit}%
 \def\sl{\fam\slfam\twlsl}%
 \def\bf{\fam\bffam\twlbf}%
 \def\mit{\fam 1}%
 \def\cal{\fam 2}%
 \textfont0=\twlrm   \scriptfont0=\tenrm   \scriptscriptfont0=\eightrm
 \textfont1=\twli    \scriptfont1=\teni    \scriptscriptfont1=\eighti
 \textfont2=\twlsy   \scriptfont2=\tensy   \scriptscriptfont2=\eightsy
 \textfont3=\tenex   \scriptfont3=\tenex     \scriptscriptfont3=\tenex
 \textfont\itfam=\twlit
 \textfont\slfam=\twlsl
 \textfont\bffam=\twlbf \scriptfont\bffam=\tenbf
   \scriptscriptfont\bffam=\eightbf
\setbox\strutbox=\hbox{\vrule height 8.5pt depth 3.5pt width 0pt}%
\def\tt{\twltt}\normalbaselines\rm}
%
%




%

\def\b{\beta}

\def\vec{\overrightarrow}

\def\cal{\Cal}

\def\ni{\noindent}

\def\abstract{\noindent{\bf Abstract : }}

\def\sec#1{\advance \nosection by 1 \nosubsect=0 \nosubsub=0
               \removelastskip\bigskip\leftline{\bf\number\nosection.\ #1}
               \medskip}
\def\subsec#1{\advance \nosubsect by 1 \nosubsub=0
               \removelastskip\medskip
               \leftline{\bf\hskip\parindent{\number\nosection.\number
               \nosubsect\ #1}}
               \medskip}
\def\subsubsec#1{\advance \nosubsub by 1
                  \removelastskip\medskip
                  {\bf\hskip\parindent{\number\nosection.\number\nosubsect.
                  \number\nosubsub\ #1}}
                  \medskip}

\def\REF{\newpage{\bf\centerline{REFERENCES}}\vskip 2cm}
\def\FIG{\newpage{\bf\centerline{FIGURE CAPTIONS}}\vskip 2cm}
\newcount\noref
\newcount\nofig
\newcount\noeq
\noeq=0
\nofig=0
\noref=0
\def\ad{\advance \noeq by 1}

\def\ref{\advance \noref by 1\item{[\number\noref]}}
\def\fig{\advance \nofig by 1\item{Fig. \number\nofig\ :}}
\hsize 16.2cm \vsize 23cm
\twlpoint
\TagsOnRight

\twlpoint \baselineskip 19pt
{\bf \centerline{COARSENING IN THE $\bold
q$-STATE POTTS MODEL AND THE}
\centerline{ISING MODEL WITH GLOBALLY CONSERVED
MAGNETIZATION}}

\vskip 2.cm

\smc\centerline{Cl\'ement Sire$^1$ and Satya N. Majumdar$^2$}\rm

\vskip 1cm\it $^1:$ Laboratoire de Physique Quantique (URA 505 du CNRS),
Universi\-t\'e Paul Sa\-ba\-tier,  31062 Tou\-louse Cedex, France (e-mail:
clement\@siberia.ups-tlse.fr).

$^2:$ Department of Physics, Yale University, New Heaven, CT 06511, USA
(e-mail: satya\@cmphys.eng.yale.edu).\rm

\vskip 1.6cm \baselineskip 19pt

\ni {\bf Abstract:} We study the nonequilibrium dynamics of the $q$-state
Potts model following  a quench from the high temperature disordered phase to
zero temperature. The time dependent two-point correlation functions of the
order parameter field satisfy dynamic scaling with a length scale $L(t)\sim
t^{1/2}$. In particular, the autocorrelation function decays as
$L(t)^{-\lambda(q)}$. We illustrate these properties by  solving exactly the
kinetic Potts model in $d=1$. We then analyze a Langevin equation of an
appropriate field theory to compute these correlation functions for general
$q$ and $d$. We establish a correspondence between the two-point correlations
of the $q$-state Potts model and those of a kinetic Ising model evolving with
a fixed magnetization $(2/q-1)$. The dynamics of this Ising model is solved
exactly in the large $q$ limit, and in the limit of a large number of
components $n$ for the order parameter. For general $q$ and in any dimension,
we introduce a Gaussian closure approximation and calculate within this
approximation the scaling functions and the exponent $\lambda (q)$. These are
in good agreement with the direct numerical simulations of the Potts model as
well as the kinetic Ising model with fixed magnetization. We also discuss  the
existing and possible experimental realizations of these models.

\vskip 0.9cm \ni PACS numbers: 05.70.L, 05.40.+j., 82.20.-w.

\ni To be published in Physical Review E (Version of\today).

\newpage \baselineskip 23.5pt {\bf\centerline{I. Introduction.}}

Coarsening of domains of equilibrium ordered phases, following a quench from
the disordered homogeneous phase to a regime where the system develops long
range order, is widely observed in many physical systems such as binary
alloys, liquid crystals, magnetic bubbles, Langmuir films and soap bubbles
[1][2]. After the quench, domains of the different ordered phases form and
grow with time as the system attains local equilibrium on larger and larger
length scales. A dynamic scaling hypothesis suggests that at late times the
system is left with a single length scale (linear size of a typical domain)
which grows with time as $L(t)\sim t^{n}$, where $n$ depends on the
conservation laws satisfied by the dynamics [1][2]. For systems with only two
types of ordered phases (such as a binary alloy or the Ising model), the
nonequilibrium coarsening dynamics have been extensively studied
experimentally, numerically and by approximate analytical methods.
Comparatively, much less is known when there are more than two ordered phases.
A particular example of the latter class of models is the $q$-state Potts
model which has $q$ ordered phases [3]. For $q=2$, this corresponds to the
Ising model, and there are experimental realizations also for $q=3,4,\infty$
[3]. As $q$ increases, the morphology of the coarsening patterns changes from
one of large connected interpenetrating domains to one of more and more
isotropic droplets. The limit $q\to\infty$ is known to correctly describe the
evolution of a dry soap froth and the growth of metallic grains [4]. The
reason for taking $q=\infty$ is to prevent bubbles represented by different
Potts indices from coalescing. In a way, a finite but large $q$ Potts model
describes a froth in which wall breakage occurs with a probability of order
$1/q$ (Fig. 1). Most studies of soap bubbles have so far focussed on
geometrical properties of the froth. For instance, mean-field treatments
[5][6] and numerical simulations [4], as well as experiments [7], have
addressed issues such as the joint distribution of bubble area and
coordination number. However, while in other physical problems of coarsening
the dynamic scaling and the correlation functions have been of central
interest, no theoretical analysis exists of spatial and temporal correlations
in a froth.

For the Potts model with finite $q$, there have been only numerical studies of
the growth law of domains, confirming $L(t)\sim t^{1/2}$ [4][8], and
substantiating the scaling of the equal-time correlation function [8].
However, to our knowledge, the two-time correlation has never been studied
previously, not even numerically.  In this paper, following our recent letter
[21], we show that the autocorrelation $A(t)=\langle {\phi}(\bold
r,0){\phi}(\bold r,t)\rangle$,  measuring the correlation of the order
parameter field ${\phi}(\bold r,t)$ with its initial value ${\phi}(\bold
r,0)$, decays as $A(t)\sim L(t)^{-\lambda(q)}$. Thus, while the growth
exponent remains $1/2$ for all $q$, the autocorrelation exponent depends
explicitly on $q$. In [21], we established a correspondence between the
dynamics of the $q$-state Potts model and that of a kinetic Ising model
evolving with a fixed magnetization $(2/q-1)$. In the present work, we explore
more of this correspondence and derive several new results.

For $q=2$, the Potts model is by definition identical to the Ising model. For
$q>2$, our correspondence  makes contact between the kinetic Potts model and a
kinetic Ising model evolving with a fixed magnetization. This is useful not
only because Ising models are conceptually much easier to think about, but
also due to the fact that they are much easier  to access experimentally,
especially for the measurement of nonequilibrium dynamical properties. For
example, in a recent experiment using video microscopy [27], the
autocorrelation exponent has been measured for a liquid crystal sample
confined between two parallel plates.  The plates were designed such that
molecules were bound to align along two perpendicular directions at the
surface of the two plates. The two directions of the ``helix'' described by
the molecules between the two plates (clockwise or anticlockwise) then
represent the two possible ``spin'' states. This system then obeys Ising
symmetry in $d=2$ and  corresponds to $q=2$  (fixed 50\%-50\% mixture of
``up'' and  ``down'' phases). The experimentally measured autocorrelation
exponent ${\lambda }_{Exp}=1.246\pm 0.079$ [27] is in good agreement with the
numerical simulation of the $d=2$ kinetic Ising model,
${\lambda}_{Num}=1.25\pm 0.01$ [9] (see also section VI).  In Ref. [9], it was
also argued heuristically that ${\lambda}=5/4$ for $d=2$.  The Gaussian
closure approximation for $q=2$ leads to ${\lambda}_{GCA}\approx 1.286$ [15]
(see also section V). In principle, for other values of $q$, $\lambda(q)$
could be measured in such a system if one succeeds in maintaining  a fixed
concentration of ``up'' phase, different from 50\%.

For $q>2$, exploiting our correspondence to an Ising model,  another possible
experimental system for measuring ${\lambda}(q)$ might be  magnetic bubbles
[26][28]. Magnetic bubbles are a natural realization of the kinetic Ising
model where increasing the magnetic field leads to the coarsening of the
system.  In principle, $\lambda(q)$ could be measured if a constant
magnetization path in the phase diagram is chosen [28].  Note that in order to
determine $\lambda(q)$, the ``real time''  scale which, as already mentioned,
is mapped onto a  function of the magnetic field in this system, is not
explicitly needed  since the definition of $\lambda(q)$ only involves the
domains length  scale, $L(t)$.

The paper is organized as follows. In section II, we solve exactly the
dynamics of the $d=1$  $q$-state Potts model and compute the equal-time and
the two-time correlation functions. We find ${\lambda}(q,d=1)=1$ for all $q$,
and show that this problem is equivalent to the $d=1$ Glauber dynamics for the
Ising model with constant magnetization $m_0=2/q-1$.  In section III, we
present an exact theory of the coarsening of this Ising model with globally
conserved magnetization $m_0=2/q-1$, in the $q\to\infty$ limit, and in any
dimension $d>1$. The distribution of droplet radii and the equal-time
correlation functions are computed exactly and we find
$\lambda(q=\infty,d)=d$. Section IV is devoted to the exact solution of an
$O(n)$ model evolving with a globally conserved magnetization, in the
$n\to\infty$ limit. The correlation functions are again calculated and
$\lambda(q,n=\infty,d)=d/2$. In section V, we first establish a general
correspondence between the dynamics of the $q$-state Potts model and that of
the Ising model evolving with a fixed magnetization $2/q-1$, and then present
a non trivial extension of the GCA [15] to this Ising model (a brief version
of which was communicated earlier [21]).  This approach gives very accurate
results for general $q$ and also reproduces the exact results obtained in the
different limits mentioned above. Finally, in section VI, we present extensive
numerical simulations for the Potts and Ising models separately, which confirm
the mentioned correspondence and show that the GCA is indeed remarkably
accurate.

As a prelude to the following sections, we first define the Potts model
Hamiltonian as [3], $$ \Cal H=-\sum_{<ij>}\delta_{\sigma_i \sigma_j}\tag I.1
$$ $\sigma_i$ takes integer values $\sigma_i=1,\hdots,q$. The sum is over
nearest neighbors, but can be extended to the next shell of neighbors, as is
needed in $T=0$ Monte Carlo simulations, in order to avoid pinning to the
lattice (see section VI).

\vskip 0.5cm {\bf\centerline{II. $\bold q$-state Potts model in one
dimension.}}

We now consider the zero temperature Glauber dynamics of the $q$-state Potts
model in one dimension. Contrary to higher dimensions where the domain growth
is driven by interfacial tension, coarsening in $d=1$ occurs via the diffusion
and annihilation of kinks (Fig. 2-a). We start with a totally random initial
condition where each of the $q$ phases is present with equal density $c=1/q$.
The zero temperature dynamics proceeds as follows: a spin is selected at
random and its value is changed to that of either of its neighbors with equal
probability. It gives rise to three possible situations as shown in Fig. 2. We
now focus on one particular phase with an arbitrary Potts index $\sigma=l$ and
define an indicator function $\phi(x,t)$ which is 1 if the site $x$ is
occupied by this $l$-th phase and 0 otherwise. In this new two-phase system,
the density of $1$'s is $c=1/q$, and the $0$'s, representing the $(q-1)$ other
phases, have a density $1-c$. For later convenience, we prefer the $\phi=0,1$
convention to the more usual spin representation $\phi=\pm 1$. The density $c$
is related to the usual magnetization $m_0$ by the relation $m_0=2c-1$. Then
the dynamics of $\phi$ is governed by the Glauber dynamics [10] of the Ising
model with a constant magnetization $m_0=2/q-1$. This is illustrated in  Fig.
2-b.

Using this mapping, the equal-time correlation function $G(r,t)=\langle
\phi(r,t) \phi(0,t)\rangle$, in the scaling limit, is given by [10], $$
qG(r,t)=\left( 1-\frac 1 q\right)\text{erfc}\left(\frac{r}{2\sqrt t}
\right)+\frac 1 q \tag II.1 $$ where $\text{erfc}(x)= {2\over {\sqrt
\pi}}\int_{x}^{\infty}\exp (-u^2)\,du$. We now point out that in the
$q\to\infty$ limit this correlation function is related to the probability
distribution of spacings between domain walls (kinks in $d=1$). The quantity
$qG(r,t)$ measures the probability that the spins at 0 and $r$ have the same
value at time $t$. For finite $q$, this is different from the probability that
they belong to the same domain. However, when $q\to\infty$, two spins which
are equal are necessarily in the same domain (bubble). Then, $qG(r,t)$ is the
probability of having a domain of length $r$ or more. It is then a standard
result that the normalized spacing distribution of kinks $P(r,t)$ is given by
$P(r,t)=2qL(t)\frac{\partial^2 G(r,t)}{\partial r^2}$. For large times, this
distribution also obeys scaling: $P(r,t)=L(t)^{-1} p(r/L(t))$, where
$p(x)=\frac\pi 2\left(\frac{q}{q-1}\right)^2 x\exp\left(-\frac \pi 4
\left(\frac{q}{q-1}\right)^2 x^2\right)$. This result, in the $q\to +\infty$
limit, was first obtained by Derrida et al. by using an analogy to a random
walk problem [11]. Interestingly, this Wigner distribution is also identical
to the eigenvalue spacing distribution for a real random matrix [12].

We also note that this result coincides with the spacing distribution in the
reaction diffusion model $A+A\to A$ [13]. This can be understood in the
following way. Representing a kink by a particle $A$ on a $d=1$ line, the
dynamics of the Potts model can be mapped to that of a reaction diffusion
model where the particles $A$ diffuse, annihilate and coagulate according to
the following rules: each particle undergoes diffusion until two of them meet
in which case they either annihilate ($A+A\to\emptyset$) with probability
$1/(q-1)$, or coagulate ($A+A\to A$) with probability $(q-2)/(q-1)$ (see Fig.
2-c).  For $q=2$, the particles only annihilate and hence, this is equivalent
to the Glauber model [10], whereas in the $q\to\infty$ limit they only
coagulate. Both these limits have been studied previously [13][14]. In the
Glauber case, the two-point correlation function has been calculated
analytically but the spacing distribution is still unknown. On the other hand,
for the $A+A\to A$ model, the spacing distribution is known exactly but there
was no analogy to any spin model and hence correlation function. Our present
result establishes that the dynamics of these two problems are two different
solvable limits of that of the $q$-state Potts model.

We now consider the two-time correlation function $C(r,t)=\langle
\delta_{\sigma(r,t), \sigma(0,0)}-1/q^2\rangle$. In terms of the indicator
field $\phi$, $C(r,t)=\langle\phi(r,t)\phi(0,0)-1/q^2\rangle$, which is
expected to scale as $C(r,t)\sim L(t)^{-\lambda} c(r/L(t))$ [9]. Thus, the
autocorrelation $A(t)=C(0,t)$ decays as $A(t)\sim L(t)^{-\lambda}$, where
$\lambda$ is a non trivial non equilibrium exponent [9]. In $d=1$ and for all
$q$, we find that $C(r,t)$ satisfies a diffusion equation and is given by
$C(r,t)\sim t^{-1/2}\exp(-r^2/2t)$. Since the length scale $L(t)\sim t^{1/2}$,
we establish the scaling of $C(r,t)$ and find $\lambda=1$ for all $q$. This is
consistent with our general exact result that $\lambda=d$ in the $q\to\infty$
limit, as we argue in the following sections.

In one dimension, we have shown that the evolution of the two-point
correlation functions in the kinetic Potts model can be exactly mapped to that
of the Ising model with fixed magnetization $(2/q-1)$. In section V, we will
show that this correspondence essentially holds  even in higher dimensions.
This fact motivates us to study the dynamics of the Ising model evolving with
fixed magnetization. In the following section, we show that this problem can
be exactly solved in the $q\to\infty$ limit.

\vskip 0.5cm {\bf\centerline{III. Small volume fraction limit.}}

In this section, we study the coarsening of a magnetic system with $globally$
conserved magnetization $m_0=(2/q-1)$, or equivalently with a density $c=1/q$
of the minority phase (up), in the $c\to 0$ limit. This becomes a modified
version (suited for globally fixed magnetization) of the celebrated
Lifshitz-Slyozov (LS) theory [16][25] which describes the coarsening of a
two-phase system with $local$ conservation (model $B$ [17]). On a discrete
lattice, both models are described by spin-exchange Kawasaki dynamics.
However, in one case the exchange occurs between spins on any two arbitrary
sites (global conservation) whereas in the other case, the exchange occurs
only between two sites which are nearest neighbor to each other (local
conservation). For $c=1/2$, the globally conserved model (hereafter called
model $A(c)$) has been shown to be in the same universality class as the
standard model $A$ dynamics, so far as the dynamical exponent and the domain
growth exponent are concerned [18][19]. In section VI, we will show that the
autocorrelation exponent $\lambda$ is also the same for a strict global
$c=1/2$ conservation and for model $A$, namely, $\lambda\approx 1.25$.
However, for $c<1/2$, we find (see section VI) that $\lambda$ is explicitly a
function of $c$. At the end of this section, we show (within the framework of
LS theory which is exact in the zero volume fraction limit) that $\lim_{c\to
0}{\lambda}(c)=d$. We recover this exact result also from our approximate
treatment of the appropriate field theory (see section V). This result is
further verified from the direct numerical simulations (section VI) of the
$q$-state Potts model (large $q$ limit) as well as the model $A(c)$.

In the $c\to 0$ limit, the coarsening pattern consists of circular bubbles
which are growing, but are always far apart from each other so that they never
coalesce. A mean-field treatment assuming no correlation between these bubbles
should then be exact in the limit $c\to 0$. In the following, we determine
$N(R,t)$, the density of bubbles with radius $r$ at time $t$.

The equation of motion for the radius $R_i(t)$ of bubble $i$ is, $$
\frac{dR_i}{dt}=-\frac{1}{R_i}+\lambda_1 \tag III.1 $$ The first term of the
right hand side, in which the surface tension dependent coefficient has been
normalized to 1, contributes to make the bubble shrink, in order to minimize
locally the interface length between the two phases. The second term,
$\lambda_1(t)$, is an effective time-dependent magnetic field, playing the
r\^ole of a Lagrange multiplier (see section V) fixing the constraint that
$\sum_i R_i^d$ must be a constant proportional to $cV$, where $V$ is the
volume. Note that this equation of motion Eq. (III.1) can also be directly
derived from the model $A$ equation of motion with this additional Lagrange
multiplier needed to keep the magnetization fixed. The density $N(R,t)$
satisfies a continuity equation of the form, $$ \frac{\partial N}{\partial
t}+\frac{\partial}{\partial R} \left(\frac{dR}{dt}N\right)\tag III.2 $$ At
late times we look for a scaling solution of the form $N(R,t)\approx
L(t)^{-(d+1)} F(R/L(t))$, where $L(t)$ is proportional to the average radius
of a growing bubble. The conservation law demands that the prefactor decays as
$L(t)^{-(d+1)}$. Inserting Eq. (III.1) and the scaling form in Eq. (III.2),
one easily sees that all terms are of the same order provided $L\frac{dL}{dt}$
is a constant, so that $L(t)\sim t^{1/2}$. Thus, in contrast to model $B$
dynamics where $L(t)\sim t^{1/3}$, the model $A(c)$ has the same growth law as
model $A$. We now set $L\frac{dL}{dt}=1$ and $\lambda_1(t)=\nu/L(t)$, where
$\nu$ is to be determined by imposing that the solution of the scaling
equation has physical limits. This form for $\lambda_1$ is justified by the
fact that, in Eq. (III.1), $\lambda_1$ scales as $R^{-1}$. This fact will be
physically justified in section V in terms of balance between the interface
and magnetic energies of a droplet. From now on, we present explicit results
for $d=2$, but the generalization to $d>2$ is straightforward. The scaling
function $F$ then satisfies the following differential equation: $$
F'(x)=\frac{1-3x^2}{x(x^2-\nu x+1)}F(x)\tag III.3 $$ The condition that $F(x)$
goes to zero for large $x$ demands that $\nu=2$, so that the function
multiplying $F(x)$ in Eq. (III.3) has a double pole (at $x=1$). Then, solving
this differential equation we get, $$ \aligned
F(x)&=A\frac{x}{(1-x)^4}\exp\left(-\frac{2}{1-x}\right), \quad\text{ for }
x\leq 1\\ &=0,\quad \text{ for } x\geq 1 \endaligned\tag III.4 $$ where the
constant $A$ is determined from the conservation condition. This scaling
function is plotted in Fig. 3 and appears to be wider than the LS form for
model $B$ (local conservation) [24].

We now compute, in the limit $c\to 0$, the equal-time correlation function
$G(r,t)=\langle \phi(r,t) \phi(0,t)\rangle/c$, where $\phi$ is the density
field as defined in the previous section. The function $G(r,t)$ can be
computed from the normalized radius distribution function
$N_0(R,t)=N(R,t)/\int \pi u^2 N(u,t)\, du=N(R,t)/cV$ in the following way. By
definition, $$ G(r,t)=\frac{ \int \phi(\bold x+\bold r)\phi(\bold x)\,d^2\bold
x} { \int \phi(\bold x)\,d^2\bold x}=\frac{1}{cV}\sum_{i,j} \int
\chi_{i}(\bold x+\bold r)\chi_{j}(\bold x)\,d^2\bold x\tag III.5 $$ where the
indices $i$ and $j$ run over all the bubbles and $\chi_i(\bold x)$ is the
characteristic function of the $i$-th bubble. In the $c\to 0$ limit, bubbles
are strictly circular and $\chi_i(\bold x)=\theta(R_i-|\bold x|)$, where
$\theta$ is the usual step function and $R_i$ is the radius of the $i$-th
bubble. In this low area fraction limit, the bubbles are far apart from each
other and for finite $r$, only the terms corresponding to $i=j$ contribute to
the sum in Eq. (III.5). Then, we obtain the exact result in the $c\to 0$
limit:  $$ G(r,t)=\int dR\, N_0(R,t)\int d^2\bold x \chi_{R}(\bold x+\bold r)
\chi_{R}(\bold x)\tag III.6 $$ The second integral is just the overlap area
between two disks with radius $R$, with their centers separated by a distance
$r$. The final expression is,  $$ G(r,t)=2\int_{r/2}^{+\infty} dR\,R^2
N_0(R,t)\left(\arccos\left(\frac{r}{2R}\right)
-\frac{r}{2R}\sqrt{1-\frac{r^2}{4R^2}}\right)\tag III.7  $$ In the scaling
limit, $N_0(R,t)\approx {1\over {cV L(t)^3}}F(R/L(t))$, where $F(x)$ is given
by Eq. (III.4), and therefore, from the above equation, we find explicitly
that $G(r,t)\approx g(R/L(t))$. Thus, dynamic scaling is established. The
function $g(x)$ with $x$ normalized such that $g(1)=1/2$ is shown in Fig. 4,
and is seen to be in nice agreement with large $q$ Potts model simulations,
and the field theory results presented below. Notice however that the
mean-field equal-time correlation function has a finite support.

In the $c\to 0$ limit, it is also simple to calculate the autocorrelation
exponent $\lambda$. Since bubbles do not coalesce in this limit and their
centers do not diffuse, the autocorrelation $A(t)\approx\langle \phi(\bold
x,t)\phi(\bold x,0)\rangle$ is essentially the survival probability of a
bubble up to time $t$. After time $t$, the number of bubbles left is
$N(t)=\int dR\, N(R,t)\sim L(t)^{-d}$, and therefore $\lambda=d$, in $d$
dimensions. This result is confirmed by numerical simulations presented in
section VI.

\vskip 0.5cm {\bf\centerline{IV. Large $\bold n$ calculation.}}

Another example where the correlation functions (both equal-time and two-time)
and the exponent $\lambda$ can be calculated in presence of a time dependent
magnetic field $\lambda_1(t)$ (to keep the average magnetization $m_0$ fixed),
is the large $n$ ($n\to \infty$) limit of the classical $O(n)$ vector model.
The $O(n)$ model is described by an $n$ component order parameter field ${\vec
\phi} (\bold r,t)=[{\phi}_1(\bold r,t),\ldots, {\phi}_n(\bold r,t)]$ and a
coarse-grained Landau-Ginzburg free energy functional, $$ \Cal H(\vec
\phi)={1\over {2}}\int d^d\,{\bold r} \left[ (\nabla {\vec \phi})^2+r_0 {\vec
\phi}^2+ {u\over {2n}}\left({\vec \phi}^2\right)^2-2\lambda_1(t)
\sum_{\alpha}{\phi}_{\alpha}\right] \tag IV.1 $$ where $\lambda_1(t)$ is a
Lagrange multiplier to keep ${1\over V}\sum_{\bold r} {\phi_\alpha} (\bold r,
t)$ fixed at $m_0=2/q-1$ where $V$ is the system size. The model $A$ equation
describing the overdamped relaxation is in general nonlinear and hard to
solve. However, in the large $n$ limit, this equation can be linearized in a
self-consistent way [20], and therefore, is exactly solvable in that limit. In
Fourier space, this linearized equation (at $T=0$) reads, $$ {{\partial
{\phi}(\bold k,t)}\over {\partial t}}=-[k^2+r_0+uS_0(t)] \phi (\bold
k,t)+\lambda_1(t){\sqrt V}{\delta}_{\bold k=0} \tag IV.2 $$ where
self-consistency for the structure factor demands that, $S_0(t)={1\over
V}\sum_{\bold k} S(\bold k,t)$, where $S(\bold k,t)=\langle\phi(\bold
k,t)\phi(-\bold k,t)\rangle$. Note that $r_0<0$, since we are in the ordered
phase. We have dropped the subscript $\alpha$ from ${\phi}_{\alpha}$ since in
the large $n$ limit, the different components of $\vec \phi$ are completely
uncorrelated with each other. The Lagrange multiplier $\lambda_1(t)$ is
determined from Eq. (IV.2), by demanding that $\phi (\bold k=0,t)=m_0{\sqrt
V}$ and is given by, $\lambda_1(t)=m_0[r_0+uS_0(t)]$. We use random initial
conditions, for which $S(\bold k,0)=\Delta-m_0^2$ for $\bold k\neq {0}$, where
$\Delta$ is of $O(1)$ and $S(\bold k=0, 0)=m_0^2V$. Writing,
$Q(t)=\int_{0}^{t}[r_0+uS_0(t')]\,dt'$, we get from Eq. (IV.2), for $\bold k
\neq 0$, $$ S(\bold k,t)=(\Delta-m_0^2)\exp [-2(k^2t+Q(t))]. \tag IV.3 $$ The
self-consistency condition now reads, $$ {dQ\over {dt}}=r_0+um_0^2+{u\over
V}\sum_{\bold k} S(\bold k,t) \tag IV.4 $$ Plugging in the expression for
$S(\bold k,t)$ from Eq. (IV.3) into Eq. (IV.4), and taking the thermodynamic
limit $V\to \infty$, we get $$ {dQ\over
{dt}}=r_0+um_0^2+u(\Delta-m_0^2){\Gamma}t^{-d/2}\exp [-2Q(t)] \tag IV.5 $$
where $\Gamma$ is a constant which depends on the dimension $d$ and the upper
cutoff $\Lambda$ of the theory. This equation is solved by making the
$ansatz$, $Q(t)=A+B\log t$. The left hand side of Eq. (IV.5) then decays as
$t^{-1}$, whereas the leading order term on the right hand side is a constant.
So, for consistency, one needs $B=-d/4$, so that the leading order term on the
right hand side is identically zero. As a consequence, the structure factor
for $\bold k\ne 0$ (Fourier transform of the equal-time correlation function)
can be written in the form, $$ \aligned S(\bold k,t)\approx L(t)^{-d}s(\bold
kL(t)),\qquad\text{ with }\\ L(t)\approx t^{1/2},\quad\text{ and }\quad
s(x)=C\exp(-2x^2) \endaligned\tag IV.6 $$ where $C$ is a constant. We thus
obtain the scaling of the correlation function with the expected domains
length scale, $L(t)\sim t^{1/2}$.

Similarly, the two-time correlation function, $C(\bold k,t)=\langle \phi
(\bold k,0)\phi (-\bold k, t) \rangle$, for $\bold k\neq 0$, evolves as
$C(\bold k,t)=(\Delta-m_0^2) \exp [-(k^2t+Q(t))]$. The autocorrelation
function, defined as $A(t)= \langle \phi (\bold r,0)\phi (\bold r,
t)\rangle-m_0^2$, then decays as $t^{-d/4}\sim L(t)^{-d/2}$. Thus, the
autocorrelation exponent is $\lambda=d/2$, as in the $m_0=0$ case [20]. This,
however, is not unexpected because the limit $n\to \infty$ decouples the
different $\bold k$ modes, and the time dependent magnetic field just couples
to $\bold k=0$ mode which has no effect on the evolution of the $\bold k\neq
0$ modes apart from modifying the prefactor. Therefore, in order to see the
dependence of $\lambda$ on $m_0$, one has to include the $O(1/n)$ corrections
which is a very hard task [2].

\vskip 0.5cm {\bf\centerline{V. Field theory and Gaussian closure
approximation.}}

In our earlier letter [21], we constructed a field theory of the $q$-state
Potts model in terms of the coarse-grained ``occupation density'' fields $\{
{\phi}_l(\bold r,t);\; l=1,2,\ldots q\}$ such that ${\phi}_l$ assumes the
value $1$ in the interior of the $l$-th ordered phase and decays continuously
to 0 outside. Consequently, inside any ``bubble'' of one phase, only one of
the ${\phi}_l$'s is close to $1$ and the others are all close to 0. We thus
require a potential with $q$ degenerate minima at $[1,0,0,\ldots 0]$,
$[0,1,0,\ldots 0]$, $\ldots$ $[0,0,0,\ldots 1]$, which prevents two different
bubbles from sharing the same position in space. A suitable free energy
functional is [21], $$ \aligned \Cal F[\lbrace \phi_l\rbrace ]=&\int d^d{\bold
r}\, \left[{\sum_{l=1}^q} \left( {1\over {2}}(\nabla
{\phi}_l)^2+V({\phi}_l)\right) -\lambda_1\left( {
\sum_{l=1}^q}\phi_l-1\right)\right.\\ & \left.+{\lambda}_2\left({
\sum_{l=1}^q}\left({\phi}_l-\frac 1 q\right)^2-{\frac{q-1}{q}}\right)^2\;
\right] \endaligned\tag V.1 $$ where $\lambda_1(\bold r,t)$ is a Lagrange
multiplier enforcing the constraint $\sum_l\phi_l=1$, and $\lambda_2$ is a
constant of $O(1)$ such that the state $[1/q,\ldots,1/q]$ is unstable, and
$V(\phi)\sim \phi^2(1-\phi)^2$ is the usual double well potential with minima
at 0 and 1. Then, the equation of motion is, $$ {{\partial {\phi}_l}\over
{\partial t}}= {\nabla}^2 {\phi}_l-
V'({\phi}_l)+{\lambda}_1-4{\lambda}_2\left({\phi}_l -\frac 1 q\right)\left[{
\sum_{l'=1}^q}\left({\phi}_{l'}-\frac 1 q \right)^2-{\frac{q-1}{q}}\right]
\tag V.2 $$ and ${\lambda}_1={1\over q}\sum_{l} V'({\phi}_{l})$ by demanding
$\sum_l\phi_l=1$ in Eq. (V.2). Note that for $q=2$ (Ising model),
${\lambda}_1=0$ by virtue of the condition ${\phi}_1+{\phi}_2=1$ and
${\lambda}_2$ can be chosen to be $0$ since this term only renormalizes $V$.
Then, one recovers the usual time-dependent Landau equation for the Ising
model. For $q>2$, we also note that this evolution equation has a form similar
to that of Eq. (2.10) of Lau $et$ $al.$ [8]. In Fig. 1, we show a late time
configuration of domains generated by Eq. (V.2) for $q=30$.

The two-point correlation function for the $q$-state Potts model is defined as
$G(12)={ \sum_{l=1}^q}\langle {\phi}_l(\bold r_1, t_1){\phi}_{l}(\bold r_2,
t_2)\rangle$ and therefore equals $q\langle {\phi}_l(\bold r_1,
t_1){\phi}_{l}(\bold r_2, t_2)\rangle$ due to the symmetry between the $q$
phases. Here, ``${12}$'' is a short-hand notation for the pair of space-time
points $(\bold r_1, t_1)$ and $(\bold r_2, t_2)$. Due to the isotropy and
translational invariance in space, the only spatial dependence of these
correlation functions is through $r=|{\bold r_1 -\bold r_2}|$. Denoting the
equal-time correlation function ($t_1=t_2=t$) by $G(r,t)$, we get from Eq.
(V.2): $$ \aligned \qquad & {1\over 2}{{\partial G}\over {\partial
t}}={\nabla}^2G-q\left \langle {\phi}_l(\bold 0,t)\left(V'({\phi}_l(\bold r,
t))-{\lambda}_1(\bold r,t)\right)\right\rangle\\ &-4{\lambda}_2q\left\langle
{\phi}_l(0,t)\left({\phi}_l(\bold r,t)-\frac 1 q\right) \left(\sum_{l'=1}^q
{{\phi}_{l'}}^2 (\bold r,t)-1\right)\right\rangle \endaligned\tag V.3 $$ Note
that the two-time correlation function satisfies a similar equation.

Our first approximation is to replace $\sum_{l'} {{\phi}_{l'}}^2$ by its
average $q\langle {{\phi}_l}^2\rangle=G(0,t)$ in the third term on the right
hand side of Eq. (V.3), which becomes exact in the $q\to \infty$ limit.
Furthermore, the scaling solution $G(r,t)=g(r/L(t))$ must satisfy $g(0)=1$, so
that we can drop the term $4\lambda_2(G(r,t)-1/q) (G(0,t)-1)$ so produced.
Thus, the third term, although important in the evolution of ${\phi}_l$ since
it provides stability to the bubbles, is not crucial in the evolution of the
correlation functions, at least in the scaling limit of the large $q$ model,
but also for $q=2$ for which this term is simply absent. The two boundary
conditions for $G(r,t)$ are: (i) As $r\to 0$, $G(r,t)\to 1$ and (ii) As $r\to
\infty$, $G(r,t)\to q\langle {\phi}_l(0,t)\rangle \langle {\phi}_l(\bold r,
t)\rangle =1/q$.

Next, using $\sum {\phi}_l(\bold r,t)=1$, we get $\langle \sum {\phi}_l (\bold
r,t) {\lambda}_1(\bold r,t)\rangle=\langle {\lambda}_1\rangle(t)$, and then
given the symmetry between the $q$ phases, we can write $\langle
{\phi}_l{\lambda}_1\rangle=(1/q)\langle {\lambda}_1\rangle=\langle
{\phi}_l\rangle \langle {\lambda}_1\rangle$. Thus, without approximation, we
replace in Eq. (V.3) ${\lambda}_1(\bold r,t)$ by its average $\langle
{\lambda}_1(\bold r,t)\rangle$, which is simply a function of time. As a
result, Eq. (V.3) reduces to an equation involving only a single field
${\phi}_l(\bold r,t)$: $$ {1\over 2}{{\partial G}\over {\partial
t}}={\nabla}^2G-q\left\langle {\phi}_l (0,t)\left( V'({\phi}_l(\bold r,
t))-\langle {\lambda}_1\rangle \right)\right \rangle \tag V.4 $$
Interestingly, Eq. (V.4) is also the evolution equation for the two-point
correlation in an Ising model evolving with fixed average magnetization
$\langle m_0\rangle=2/q-1$ (equivalently, with a density of minority ``up''
spins fixed at $1/q$). The droplet domains of the minority phase in this Ising
model would correspond to the bubbles of a particular phase in the Potts
model, with the majority phase corresponding to the remaining $(q-1)$ phases.
$\langle {\lambda}_1\rangle$ acts as a time-dependent magnetic field which
prevents the minority phase from disappearing at $T=0$, and keeps the
magnetization constant.

Thus, from now on, instead of the original Potts model, we consider the
coarsening in the Ising model with globally conserved magnetization through a
time-dependent magnetic field (Lagrange multiplier). The magnetization $m_0$
is related to the value of $q$ of the Potts model via $m_0=2/q-1$.

The calculation of the two-point correlation function of this problem can be
performed approximately by extending the Gaussian closure scheme as developed
by Mazenko [15] for the usual Ising model where the magnetization remains
fixed at $m_0=0$. The essence of this approximation scheme is to invoke an
auxiliary field $m(\bold r,t)$ which is related to the order parameter field
$\phi(\bold r,t)$ via a non linear transformation $\phi(\bold r,t)=\sigma
(m(\bold r,t))$. The idea is to find a field $m(\bold r,t)$ which varies
smoothly across the interfaces as opposed to the original field $\phi(\bold
r,t)$ which changes abruptly from nearly 0 to nearly 1 across an interface.
So, the non trivial part of the scheme is to choose the appropriate mapping
function $\sigma$. In the case of the Ising model with zero magnetization,
Mazenko argued that the function $\sigma$ should be chosen as the equilibrium
profile of $\phi$ near an interface, which is determined by the solution of $$
{{d^2\sigma}\over {dm^2}}=V'(\sigma)\tag V.5  $$ with the boundary conditions
$\sigma(m)\to 1$ as $m\to +\infty$ and $\sigma(m)\to 0$ as $m\to -\infty$. The
solution is $\sigma (m)=[1+\tanh(\beta m)]/2$, which at late times can be
replaced by a step function, since domains grow (with $L(t)\sim t^{1/2}$)
whereas the interface width ($\sim \b^{-1}$ related to the coupling constant
in front of $V$) remains bounded. The auxiliary field $m(\bold r,t)$ can be
interpreted as the distance from the nearest interface. The next part of the
approximation is to assume that $m(\bold r,t)$, being smooth across the
interface, has a Gaussian distribution. The virtue of this ``minimal''
approximation is that it facilitates analytical calculation yielding non
trivial results for the correlation functions and the exponents which are in
good agreement with simulations, at least in the nonconserved case.

In order to extend this approximation to our problem with globally conserved
magnetization, we start off with the assumption that there exists, as in the
nonconserved Ising case, a nonlinear transformation $\phi(\bold
r,t)=\sigma(m(\bold r,t),t)$ with a Gaussian auxiliary field. However, several
important modifications need to be done in carrying out this extension from
the simple Ising case. First, $\langle \phi(\bold r,t) \rangle$ must be
strictly fixed at  $1/q$ at all times (by virtue of the time-dependent field
$\langle{\lambda}_1\rangle(t)$), as opposed to the Ising case where, for a
critical quench, $\langle \phi \rangle =1/2$ automatically. This also
necessarily implies that the mean of the distribution of $m$ is nonzero. The
first and second moments of the Gaussian distribution, $\langle m(\bold
r,t)\rangle=\bar m(t)$, $\langle \lbrace m(\bold r, t)-\bar m(t)\rbrace
^2\rangle=C_0(t)$ are space independent due to translational invariance. The
complete correlation function $\langle \lbrace m(\bold r_1,t_1)- \bar m
(t_1)\rbrace \lbrace m(\bold r_2, t_2)-\bar m(t_2)\rbrace \rangle=C(12)$ must
be determined self-consistently as is $C(12)$ in the Ising case [15]. Now,
from the condition $\langle \phi(\bold r,t)\rangle=1/q$, we get ${1\over
{\sqrt {2\pi C_0}}}\int_{-\infty}^{\infty}{\sigma (m)}\exp [-(m-\bar
m)^2/{2C_0}]\,dm=1/q$. Replacing $\sigma (m,t)$ at late times by the step
function $\theta (m)$ (which is $1$ for $m>0$ and $0$ for $m<0$), and thereby
neglecting terms that are of lower order in $t$, we get $\bar m(t)=-{\sqrt
{2C_0(t)}} \text{erfc}^{-1}(2/q)$. Note that for $q=2$, $\bar m=0$ as
expected. For later convenience, let us also define the correlation function
$f(12)=C(12)/{\sqrt {C_0(1)C_0(2)}}$ and denote it by $f(\bold r,t)$ when
$t_1=t_2=t$. Note that $f(0,t)=1$, and $f\to 0$ as $r\to \infty$. Even for
$t_1=t_2$, we well keep on writing $C_0(1)$ and $C_0(2)$ explicitly, although
this two numbers are equal, since we will use formal derivatives with respect
to $C_0(1)$.

The second important difference from the simple Ising case is the choice of
the mapping function $\sigma(m(\bold r,t),t)$. The explicit time dependence
introduced via $\langle {\lambda}_1\rangle$ modifies the local equilibrium
profile, thereby precluding the choice of the stationary profile
$[1+\tanh(\beta m)]/2$. In fact, since the mean of the field $m(\bold r,t)$ is
time-dependent (through $C_0(t)$ which scales linearly with $t$ as we argue
below), one can expect to get a ``sigmoid'' shaped solution only in a moving
frame, with a velocity suitably determined to neutralize the time-dependence
introduced via $\langle {\lambda}_1\rangle$. Thus, making the transformation,
${\bold r'}={\bold r} +a(t){\hat {\bold n}}$ where ${\hat{\bold n}}$ is an
arbitrary unit vector and $a(t)$ is to be determined, and demanding an
equilibrium solution, $i.e.$, ${{\partial \phi}\over {\partial t}}=0$ to
leading order in time, we find the appropriate equation for $\sigma (m,t)$:
$$ {{d^2\sigma}\over {dm^2}}+{da\over {dt}}\,{{d\sigma}\over {dm}}=
V'(\sigma)-\langle {\lambda}_1\rangle \tag V.6 $$ We now fix $a(t)$ from the
condition that the average value on both sides of Eq. (V.6) should be
identical. The average on the right hand side is zero by definition of
$\langle {\lambda}_1\rangle$. Then from Eq. (V.6), we get, $$
\frac{da}{dt}=-\frac{\langle \sigma''(m)\rangle}{\langle \sigma'(m)\rangle}
=-\frac{\int^{+\infty}_{-\infty}du\,\exp-\frac{(u-\bar m)^2}{2C_0}\sigma''(u)}
{\int^{+\infty}_{-\infty}du\,\exp-\frac{(u-\bar m)^2}{2C_0}\sigma'(u)} \tag
V.7 $$ Now $\langle{{d^2\sigma}\over {dm^2}}\rangle/\langle {{d\sigma}\over
{dm}}\rangle$ is calculated replacing $\sigma (m,t)$ by $\theta (m)$ at late
times, so that $\sigma'(m)\approx \delta(m)$ and $\sigma''(m)\approx
\delta'(m)$. This yields, $$ {{da}\over {dt}}\approx -\frac{d}{d\bar
m}\log\left(\exp-\frac{\bar m^2}{2C_0} \right)=\frac{\bar m}{C_0}\tag V.8 $$
Anticipating a scaling solution, we find that  $a(t)\sim\sqrt{C_0(t)}\sim
L(t)$, which is expected, since $L(t)$ is physically the only remaining length
scale at late times. From Eq. (V.8) and the expression of $\bar m$ as a
function of $C_0(t)$, we see that $\langle {\lambda}_1\rangle\sim L(t)^{-1}$,
which can be understood on physical grounds: local equilibrium of a bubble and
its interface requires that the surface tension energy $E_S\sim L(t)^{d-1}$
should balance the magnetic energy $E_M\sim \langle {\lambda}_1\rangle L(t)^d$
(see also section III).

Using the fact that $m(\bold r,t)$ has a Gaussian distribution, the
correlation function $G(\bold r,t)$ is given by,  $$ \aligned &\qquad\qquad
\qquad\quad G(\bold r,t)=\frac{q\gamma}{2\pi\sqrt{C_0(1)C_0(2)}}\int
dx_1\,dx_2\,\sigma(x_1)\sigma(x_2)\times\\ &\exp\left[
-\frac{\gamma^2}{2}\left(\frac{(x_1-\bar m_1)^2}{C_0(1)}+ \frac{(x_2-\bar
m_2)^2}{C_0(2)}+2(x_1-\bar m_1)(x_2-\bar m_2)\frac
{f}{\sqrt{C_0(1)C_0(2)}}\right)\right] \endaligned\tag V.9 $$ where
$\gamma=1/\sqrt{1-f^2}$ and we recall that $f(12)=C(12)/{\sqrt
{C_0(1)C_0(2)}}$ and that the arguments ``1'' and ``2'' denote $(\bold r,t)$
and $(0,t)$. The derivative with respect to $m$ involved when inserting Eq.
(V.6) in Eq. (V.4) are more easily expressed in the Fourier space associated
to the variable $x$. $G(\bold r,t)$ then takes the form, $$ \aligned G(\bold
r,t)=\frac{q}{4\pi^2}\int dk_1 &\,dk_2\,
\hat\sigma(k_1)\hat\sigma(k_2)\times\\ \exp\left[-\frac{k_1^2}{2}C_0(1)-
\frac{k_2^2}{2}C_0(2)-k_1 \right. & \left. k_2C(12) +ik_1\bar m_1+ik_2\bar
m_2\right] \endaligned\tag V.10 $$ where $\hat\sigma$ is the Fourier transform
of $\sigma$. From this expression one finds that, $$ \aligned
\left\langle\sigma(2){{d^2\sigma}\over {dm^2}}(1)\right\rangle
&=2\frac{\partial G} {\partial C_0(1)}_{|\bar m_1} \\
\left\langle\sigma(2){{d\sigma}\over {dm}}(1)\right\rangle &=\frac{\partial G}
{\partial \bar m_1}_{|C_0(1)} \endaligned\tag V.11 $$ Using this result and
noting that for large time $\frac{\bar m_1}{C_0(1)} = 2\frac{\partial\bar
m_1}{\partial C_0(1)}$, the second term on the right hand side of Eq. (V.4)
can be written as, $$ \aligned
\qquad\qquad\left\langle\sigma(2){{d^2\sigma}\over {dm^2}}(1)\right\rangle+
\frac{\bar m_1}{C_0(1)}& \left\langle\sigma(2){{d\sigma}\over
{dm}}(1)\right\rangle= \\ \qquad\qquad 2\left(\frac{\partial G}{\partial
C_0(1)}_{|\bar m_1}+\frac{\partial\bar m_1}{\partial C_0(1)}\frac{\partial
G}{\partial \bar m_1}_{|C_0(1)}\right)&= 2\frac{\partial G}{\partial
C_0(1)}=-\frac{f}{ C_0(1)}\frac{\partial G}{\partial f} \endaligned\tag V.12
$$ Therefore, the evolution equation for the correlation function can be
expressed as, $$ {1\over 2}{{\partial G}\over {\partial t}}={\nabla}^2G
+{1\over {C_0(t)}}Q(f) \tag V.13 $$ where $Q(f)=f{{\partial G }\over {\partial
f}}$. Interestingly, we notice that the last equation is identical in form to
that of the Mazenko equation [15][21] for the critical Ising case with the
exception that $G(f)$ has different expressions in the two cases. However,
this seems accidental because in our problem we need to invoke a moving frame
and therefore a different profile function satisfying Eq. (V.6). A naive
application of the Mazenko theory with the mapping function determined by the
equilibrium profile Eq. (V.5) leads to inconsistent and unphysical results.
For $q=2$ ($c=1/2$, Ising critical), the velocity of the moving frame
$\frac{da}{dt}\approx \bar m/C_0$ is zero identically, and our expression then
reduces to the Mazenko result.

Replacing $\sigma$ by the $\theta$ function in Eq. (V.9), we get the leading
term for $G(f)$ for large time: $$ G(f)={q\over {\sqrt
\pi}}\int_{0}^{\infty}dy\, {\exp\left[- \left(y+p{\sqrt
{2\over{1+f}}}\right)^2\right]}\text{erf}\left[{\sqrt
{{1+f}\over{1-f}}}y\right] \tag V.14 $$ where $p=\text{erfc}^{-1}(2/q)$. In
principle, the function $G(f)$ can be inverted, so that $Q$ is implicitly a
function of $G$. Note that this function has the correct short (as $f\to 1$,
$G(f)\to 1$) and long distance (as $f\to 0$, $G(f)\to 1/q$) behaviors. In
addition, for $q=2$, it reduces to the Ising case [15], $G(f)={2\over
{\pi}}\tan^{-1} [\sqrt {(1+f)/(1-f)}]$. We now substitute the scaling form
$G(r,t)=g(r/L(t))$ in Eq. (V.13). A scaling solution is obtained provided
$C_0(t)\sim L(t)^2\sim t$, which leads to the expected form for $L(t)$. The
fact that $C_0(t)$ is proportional to $L^2(t)$ is consistent with the
definition of $C_0(t)$ as a two-point correlation function of the field $m$,
which has the physical meaning of a distance. The condition $L^2(t)\sim t$ is
obtained by demanding that all terms in Eq. (V.13) obtained by plugging in a
scaling form for the correlation function are of the same order as a function
of time. More precisely, setting $C_0(t)\approx 4t/{\mu}$ and $x=r/L(t)$, we
get from Eq. (V.13), $$ {{d^2g}\over {dx^2}}+\left({{d-1}\over {x}}+x\right)
{{dg}\over {dx}}+{\mu}Q(g)=0 \tag V.15 $$ which defines a closed eigenvalue
equation for the scaling function $g$. The eigenvalue $\mu$ has to be
determined by matching the short and long distance behaviors of $g(x)$. The
autocorrelation exponent $\lambda$ is then related to $\mu$ via the relation
$\lambda=d-\mu/2$, following an argument due to Mazenko [15] that we adapt to
our problem below.

Following the same line of arguments as used to derive the evolution equation
for $G$, we find that the two-time correlation function $C(\bold
r,t)=q\langle\phi(\bold r,t)\phi(0,0)\rangle-q^{-1}$ satisfies the equation,
$$ {{\partial C}\over {\partial t}}={\nabla}^2C +{1\over {C_0(r,t)}}\hat Q(f)
\tag V.16 $$ where $\hat Q(f)=f{{\partial C }\over {\partial f}}$, and $C(f)$
has the same $f$ dependence as $G(f)$. Since the two-time correlation function
decays with time, its value is very small at late times and then $Q\sim C$.
With $C_0(t)\approx 4t/\mu$, this linear equation can be solved, and $C(\bold
r,t)\sim t^{-(d/-\mu/2)/2}\exp(-\bold r^2/2t)$. Therefore, the autocorrelation
$A(t) = C(0,t)\sim t^{-(d-\mu/2)/2} \sim L(t)^{-(d-\mu/2)}$, and
$\lambda=d-\mu/2$.

In the $q\to \infty$ limit, it is possible to solve Eq. (V.15) analytically.
Neglecting terms of $O(1/p^4)$ and using $\text{erfc}(p)\sim {\exp
(-p^2)/{p\sqrt \pi}}$ for large $p$, we find $g(f)\approx {{1+f}\over
2}\text{erfc}\left[p\sqrt {{1-f}\over{1+f}}\right]$. Note that as $f\to 0$,
$g(f)=1/q+ f(1+2p^2)/q +O(f^2)$, and as $f\to 1$, $g(f)\to {1-p\sqrt
{2(1-f)/{\pi}}}$. Then, expressing the function $Q$ in terms of $g$ itself, we
find essentially three regimes. As $g\to 1/q$ (large distance), $Q(g)\approx
g-1/q$, and as $g\to 1$ (short distance), $Q(g)\approx {{p^2}\over
{{\pi}(1-g)}}$. In the intermediate regime, $g^{*}\ll g\ll 1$ where
$g^{*}\sim{{\log (q)}/q}$, $Q(g)\approx {p^2g}/2$. Note that, as $q$ becomes
larger and larger, the window over which $Q(g)$ behaves as $g-1/q$ becomes
smaller and smaller and for a large range of distance one has $f\approx 1$.
First consider the small $x$ behavior of Eq. (V.15). Using $Q(g)\sim
{p^2}/{\pi (1-g)}$, we find that $g(x) \to 1-p{\sqrt {{\mu}\over{\pi
(d-1)}}}x$, reflecting the presence of sharp interfaces. This reproduces
Porod's law [22], namely that the structure factor scaling function $F(y)$,
the Fourier transform of $g(x)$, decays as $y^{-(d+1)}$ for large argument
$y$.

For very large $q$, since in the interesting range of distance one has
$f(r,t)\approx 1$, one can find a very simple parametrization for $g$ and
$Q(g)$. Using the new variable $u\approx p\sqrt{\frac{1-f}{2}}$, we obtain, $$
\aligned g(u) &=\text{erfc}(u)\\ Q(u) &=p^2 R(u)=\frac{p^2}{2\sqrt\pi
u}\exp-u^2 \endaligned\tag V.17 $$ Because of the overall factor $p^2$ in the
expression of $Q(u)$, and since $p$ grows with $q$ $(p^2\approx \log (q))$, we
expect $\nu=\lim_{q\to\infty}\mu(q) p^2$ to be finite. Then, using the
parametrization of Eq. (V.17), $u(x)$ satisfies the following eigenequation:
$$ u''-2uu'^2+u'\left(\frac{d-1}{x}+x\right)-\frac{\nu}{4u}=0\tag V.18 $$
Porod's law gives $u\approx\sqrt{\frac{\nu}{2(d-1)}}x$ for $x\to 0$, and the
large $x$ limit is easily found to be $u\approx x/\sqrt 2$. By matching the
two regimes, one can check that $u(x)=x/\sqrt 2$ is a solution of Eq. (V.15),
provided the eigenvalue $\nu$ satisfies $\nu=2(d-1)$. The eigenfunction is
then $g(x)=\text{erfc}(x/\sqrt 2)$. Moreover, for large $q$, $\mu\sim
2(d-1)/p^2$, where $p^2\approx\log(q)$, and therefore $\lambda\approx
d-(d-1)/{\log q}$.

We note that, for $d=1$, the eigenvalue problem can be solved directly for any
$q$. The small $x$ behavior of $g(x)$ obtained from Eq. (V.15) implies
$\mu=0$. The scaling function then satisfies the differential equation
$g''+xg'=0$. As a consequence, $g(x)$ coincides with the exact solution (Eq.
(II.1)) of the $d=1$ $q$-state kinetic Potts model presented in section II.
$\mu=0$ leads to $\lambda=1$, which was also a result of the exact $d=1$
calculation. Thus, the Gaussian closure approximation is exact in $d=1$.

\vskip 0.5cm {\bf\centerline{VI. Numerical simulations.}}

We now compare our results with the direct $T=0$ simulation of the $q$-state
Potts model. We have also simulated directly our field theory (Eq. (V.2)) and
found that it evolves in a similar way as the Potts model (see Fig. 1) with
domains growing as $L^2(t)\sim t$. However, in the field theory, one needs as
many fields (4 bits real) as Potts indices which limits the maximum lattice
size ($\sim 120\times 120$), $q$ ($q_{max}\sim 50$) and the number of time
steps. The determination of $\lambda$ requires large lattice (especially for
large $q$) and large numbers of MC steps (typically $10^4$ or more) which is
easier to achieve in the Potts model simulation. The calculations have been
carried out at $T=0$ on a $800\times 800$ square lattice with equal coupling
to nearest and next nearest neighbors (NNN). NNN interactions are needed at
$T=0$ to avoid pinning to the lattice for $q>2$. It also ensures a better
isotropy of surface tension and thus of bubbles. We optimized the MC procedure
by only selecting surface Potts spins for updating, since only these can be
flipped at $T=0$. For $q\leq 20$, the results for 30 to 40 samples were
averaged (more than in [21]), whereas 10 to 20 samples were found to be
sufficient for $q>20$, due to smaller fluctuations for $L(t)$ and $A(t)$ with
increasing $q$, yielding, to our knowledge, the most extensive simulations to
date.

We found $L^2(t)\sim t$ for all $q$, confirming a result already obtained in
previous studies [4][8]. We also observed good scaling of the correlation
functions. In Fig. 4, we compare the equal-time correlation function for
$q\to\infty$ as given by the Potts model simulation, the direct simulation of
the field theory for $q=50$ fields on a $120\times 120$ lattice, the
mean-field theory of section III, and by our approximate theory, and find good
agreement between the numerical results and the two theoretical ones. For soap
bubbles, $g(x)$ measures the probability that the point $x$ belongs to the
same bubble as the origin. Since bubbles are essentially isotropic, we expect
$g$ to be closely related to the distribution of radii. Since we find that
$g(x)$ has a Gaussian tail, the distribution of areas ($A\sim x^2$) has a
Poissonian tail. This result is consistent with a maximum entropy [23] and a
mean-field [5][6] calculation. More precisely, the second derivative of $g(x)$
is proportional to the interface spacing distribution on a linear cut through
the froth, which has a Wigner form [12] in our case.

A more interesting test of our theory concerns the computation of the
autocorrelation exponent $\lambda$. The mean-field result of section III
predicts $\lambda=d$, whereas the large $n$ calculation (section IV)  leads to
$\lambda=d/2$. The more sophisticated Gaussian closure approximation of
section V  gives less trivial and more accurate results: in Tab. 1, we present
the values of $\lambda$ generated in Potts simulations, and compare them to
those obtained from the (numerical) solution of the one-dimensional eigenvalue
problem of Eq. (V.15). We find a reasonably good agreement. Notice that for
large $q$ ($q>100$) the exponent $\lambda$ obtained from Monte Carlo
simulations is probably slightly overestimated. This is due to the fact that
for a finite lattice $q=\infty$ is actually realized for a finite value of
$q$, so that one can expect that the effective number of different phases is
$q_{eff}>q$. Finite size scaling indeed confirms this fact, although
$\lambda(N)$ does not seem to have a simple form. In fact, for the values of
$q$ presented here, the finite size correction is comparable to the
statistical error bars. For instance, for $q=200$ and a $300\times 300$
lattice, we find $\lambda=1.84\pm0.02$ instead of $\lambda=1.82\pm0.02$ for a
$800\times 800$ lattice. We also note that $\lambda$ from the simulation
saturates very slowly to its $q\to\infty$ value, as predicted by our
asymptotic result. For soap bubbles, $q=\infty$ and $\lambda=2$ ($\lambda=d$
in dimension $d$). Indeed, the choice $q=\infty$ eliminates the coalescence of
bubbles with identical index, so that $A(t)\sim N(t)^{-1}\sim L(t)^{-d}$,
where $N(t)$ is the number of remaining bubbles at time $t$ (see also section
III).

We also tested the correspondence between the two-point correlation
func\-tions of the $q$-state Potts model and that of the Ising model with a
fixed average magnetization $\langle m_0\rangle= 2/q-1$, as suggested by our
analysis of section V. The dynamics is described by an infinite range Kawasaki
dynamics where two randomly selected spins are exchanged with probability 1 if
the energy is lowered, and with probability 1/2 if the energy is not changed
[18]. Again, the algorithm is optimized by keeping track of the movable
(surface) spins and by rescaling properly the unit of time. This last aspect
is not important for the determination of $\lambda$ which does not involve the
real time explicitly, but only $L(t)$. For $q>2$, this simulation appears to
be much more delicate than for the $q$-state Potts model or for the Ising
model with $q=2$. Indeed, the choice of the initial conditions is crucial as
it is already known [24] in simulations of the dynamics following an
off-critical quench in the conserved order parameter (model $B$) case. For
instance, one can start from an assembly of preformed circular bubbles with a
distribution of radius given by the mean-field expression of Eq. (III.4). The
bubbles are then randomly placed on the lattice. Although attractive, this
procedure leads to very long transient times since the correct correlations
between bubbles are long to establish through merging of bubbles. In other
words, the (absence of) correlations introduced in the initial state are very
long to destroy, a phenomenon which is amplified for large $q$.  We thus
decided to start from a more intrinsic completely random initial condition
where $N^2/q$ up spins are randomly distributed on the lattice, and considered
$800\times 800$ square lattices, with equal nearest and next nearest neighbor
couplings.  For each value of $q$, the results have been averaged over 80
samples. These large sizes are necessary in order to allow large coarsening
times. In fact, notice that the smaller is  $q$, the shorter are the
accessible coarsening times. Indeed, for a small concentration of minority
phase (as for $q=100,200$), a too small number of droplets is obtained after a
rather short time.

We found that the scaling function $g$, once normalized such that $g(0)=1$, is
only very weakly $q$ dependent, and is almost indistinguishable from the
curves already presented on Fig. 4. More interestingly, we computed the
autocorrelation $A(t)=\langle s_i(t)s_i(0)\rangle-m_0^2\sim L(t)^{-\lambda}$.
In order to determine $\lambda$ properly, and despite the long accessible
coarsening times, we also had to use an interpolation scheme introduced in [9]
for the usual Ising model. The authors argued that the effective exponent
$\lambda(t)$ measured at time $t$ should behave as,  $$
\lambda(t)-\lambda_\infty=-\frac{\log\left(\frac{A(bt)}{A(t)}\right)}
{\log\left(\frac{L(bt)}{L(t)}\right)}-\lambda_\infty\sim L(t)^{-1} \tag VI.1
$$ where $b$ is a time scaling factor chosen in the range $10-40$ depending on
$q$ and the speed of the simulation. $\lambda_\infty$ is the exponent to be
found. This relation was fairly well obtained for all $q$ and typically
modified the naive value of $\lambda$ (measured at large time) by $0.03-0.06$
depending on $q$. We insist again on the fact that,  due to the observed
sensitivity to initial conditions (totally random, preformed bubbles...), it
is possible that systematic errors are actually comparable or even bigger than
the error bars [24]. The results of these simulations are presented on Tab. 1.
Although for certain values of $q$ the difference between the obtained values
of $\lambda$ for the Potts model and the globally conserved Ising model is
bigger than the error bars (in fact only for $q=50$),  the two exponents
remain very close for all $q$. Actually, except for $q=200$, we systematically
have $\lambda_{Potts}\geq \lambda_{Ising}$ by a typical amount of 0.02.  We
cannot conclude whether this tendency is real or simply results from
systematic errors in the simulation and the various extrapolation schemes
used. However, considering the already mentioned problems affecting the Ising
simulation, but also the Potts simulation for large $q$, we cannot exclude
that these exponents are strictly identical, as suggested by our theoretical
analysis. Also notice that for $q=2$, the exact global conservation of the
magnetization does not affect the value of $\lambda$, which is consistent with
the results in [18][19].

\vskip 0.5cm {\bf\centerline{VII. Conclusion.}}

In this paper, we have studied in detail the phase ordering process following
a temperature quench in systems possessing, in general, $q$ ($q\ge 2$)
degenerate ordered phases at low temperature. We studied the dynamics of the
$q$-state Potts model which accurately describes these systems. We have
derived several exact results in different limits and obtained an important
and interesting correspondence between the two-point correlation functions of
the $q$-state Potts model and that of an Ising model evolving with fixed
magnetization $m_0=2/q-1$. This analogy has been particularly useful in
extending the Gaussian closure approximation developed for $q=2$ to the case
when $q>2$, and the results obtained from this approximation agree very well
with our direct numerical simulations. A note about GCA is worth mentioning at
this point. It is well known [2] that  up to now, the GCA fails to capture the
essential dynamics in many situations such as the model $B$ dynamics where the
order parameter is locally conserved or even model $A$ with long-range
interactions [2]. However, for short range model $A$ systems, such as
nonconserved Ising model, GCA has given reasonably good answer especially for
the autocorrelation exponent $\lambda$. So, it is not surprising that for our
system, which is also a short range model $A$ system, it produces reasonably
accurate values for the exponents ${\lambda}(q)$.

As mentioned in the introduction, the correspondence to the Ising model has
also an interesting experimental significance. We hope that this study will
motivate future experimental work to measure quantities such as $\lambda(q)$,
as it has already been done for $q=2$.

\vskip 0.5cm \noindent\it Acknowledgments: \rm We are grateful to D. Huse and
M. Seul for stimulating discussions. We thank AT\&T Bell Labs crew for its
nice support during the two past years.

\newpage \REF

\item{[1]} For a general review, see Gunton J.D., San Miguel M., Sahni P.S.,
in \it Phase Transitions and Critical Phenomena\rm, Ed. C. Domb and
J.L.Lebowitz (Academic, NY 1989), Vol. 8, p. 269; J.S. Langer, in \it Solids
Far from Equilibrium\rm, ed. C. Godr\`eche (Cambridge Un. Press, 1992).
\item{[2]} A.J. Bray, NATO ASI on \it Phase Transitions and Relaxation in
Systems with Competing Energy Scales\rm, Geilo, Norway (1993). For an even
more complete review on recent theoretical advances, see A.J. Bray, to be
published in Advances in Physics (1994). \item{[3]} F.Y Wu, Rev. Mod. Phys.
{\bf 54}, 235 (1982). \item{[4]} Grest G.S., Srolovitz D.J., Anderson M.P.,
Phys. Rev. B {\bf 38}, 4752 (1988); Anderson M.P., Grest G.S., Srolovitz D.J.,
Philos. Mag. B {\bf 59}, 293 (1989). Glazier J.A., Anderson M.P., Grest G.S.,
Philos. Mag. B {\bf 62}, 615 (1990). Note that an exhaustive list of available
experimental and theoretical results can be found in Glazier J.A., PhD Thesis
(University of Chicago, 1989), unpublished. \item{[5]} H. Flyvberg, Phys. Rev.
E {\bf 47}, 4037 (1993); see also H. Flyvberg, C. Jeppesen, Physica Scripta
{\bf T38}, 49 (1991). \item{[6]} J. Stavans, E. Domany, D. Mukamel, Europhys.
Lett. {\bf 15}, 479 (1991). \item{[7]} Glazier J.A., Gross S.P., Stavans J.,
Phys. Rev. A {\bf 36}, 306 (1987); Stavans J., Glazier J.A., Phys. Rev. Lett.
{\bf 62}, 1318 (1989); Glazier J.A., Stavans J., Phys. Rev. A {\bf 40}, 7398
(1989). \item{[8]} M. Lau, C. Dasgupta, O.T. Valls, Phys. Rev. B {\bf 38},
9024 (1988). \item{[9]} D.S. Fisher, D.A. Huse, Phys. Rev. B {\bf 38}, 373
(1988). \item{[10]} R.J. Glauber, J. of Math. Phys. {\bf 4}, 294 (1963).
\item{[11]} B. Derrida, C. Godr\`eche, I. Yekutieli, Phys. Rev. A {\bf 44},
6241 (1991). \item{[12]} For instance, see O. Bohigas, M.J. Giannoni, C.
Schmidt, \it Quantum Chaos and Statistical Nuclear Physics\rm, T.H. Seligman
and N. Nishioka eds., Lecture Notes in Physics, Vol. 263 (Springer Berlin
1986) p. 18. \item{[13]} D. ben-Avraham, M.A. Burschka, C.R. Doering, J. Stat.
Phys. {\bf 60}, 695 (1990). \item{[14]} A.J. Bray, J. Phys. {\bf A22}, L67
(1990); J.G. Amar, F. Family, Phys. Rev. A {\bf 41}, 3258 (1990). For a
complete discussion of $d=1$ coarsening models, see S.N. Majumdar and D.A.
Huse, unpublished. \item{[15]} G.F. Mazenko, Phys. Rev. B {\bf 42}, 4487
(1990); F. Liu, G.F. Mazenko, Phys. Rev. B {\bf 44}, 9185 (1991). \item{[16]}
I.M. Lifshitz, V.V. Slyozov, Sov. Phys. JETP {\bf 8}, 331 (1959); Sov. Phys.
Solid State {\bf 1}, 1285 (1960). J. Phys. Chem. Solids {\bf 19}, 35 (1961).
\item{[17]} P.C. Hohenberg, B.I. Halperin, Rev. Mod. Phys. {\bf 49}, 435
(1977). \item{[18]} F.F. Annett, J.R. Banavar, Phys. Rev. Lett. {\bf 68}, 2941
(1992). \item{[19]} A.J. Bray, Phys. Rev. Lett. {\bf 66}, 2048 (1991); and
references therein. \item{[20]} A. Coniglio, M. Zannetti, Europhys. Lett. {\bf
10}, 575 (1989). \item{[21]} C. Sire, S.N. Majumdar, Phys. Rev. Lett., in
press (1995).  \item{[22]} G. Porod, Kolloid Z. {\bf 124}, 83 (1951); {\bf
125}, 51 (1952). \item{[23]} Weaire D., Rivier N., Contemp. Phys. {\bf 25}, 59
(1984); and ref. therein. \item{[24]} T.M. Rogers, R.C. Desai, Phys. Rev. B
{\bf 39}, 11956 (1989). \item{[25]} For corrections to the LS theory, see M.
Marder, Phys. Rev. A {\bf 36}, 858 (1987), and references therein. \item{[26]}
K.L. Babcock, R.M. Westervelt, Phys. Rev. A {\bf 40}, 2022 (1989); K.L.
Babcock, R. Seshadri, R.M. Westervelt, Phys. Rev. A {\bf 41}, 1952 (1990).
\item{[27]} M. Mason, A.N. Pargellis, B. Yurke, Phys. Rev. Lett. {\bf 70}, 190
(1993). \item{[28]} M. Seul, private communication.

\newpage \FIG \item{Fig. 1:} Late time configuration for $q=30$ fields
evolving according to Eq. (V.2). Note the bubble marked by an arrow, which is
highly non isotropic, resulting from the coalescence of two bubbles with the
same index. \item{Fig. 2:} Potts model in one dimension: (a) the three
elementary dynamical flips of the central Potts spin with index $a$; (b)
associating 1's to the $a$-phase and 0's to the other phases, one obtains an
effective Glauber dynamics; (c) associating a particle $A$ to each interface,
the model is mapped on a reaction diffusion model as described in the text.
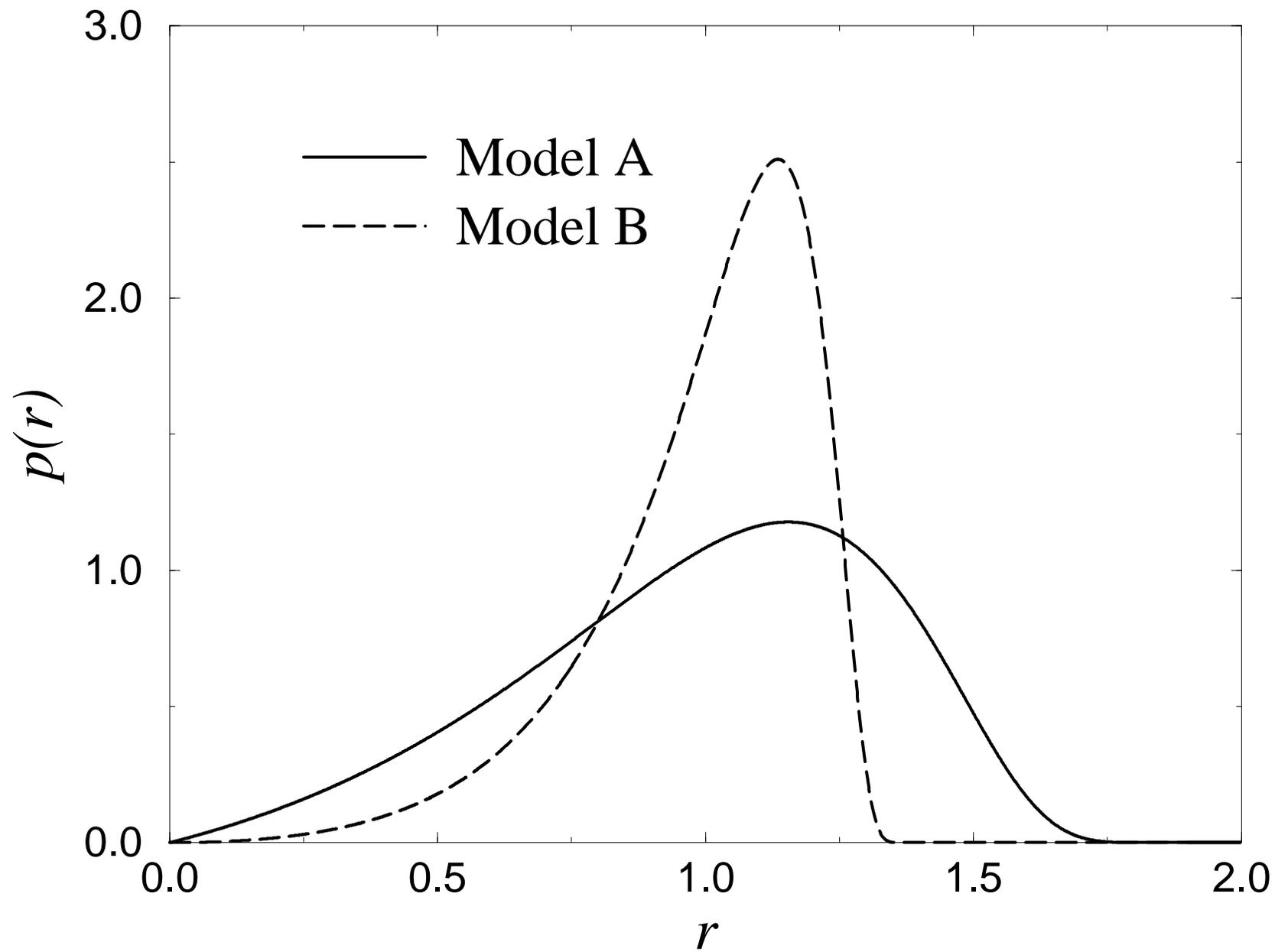
\item{Fig. 3:} Distribution of bubble radii for $d=2$ and model $A$, in the
mean-field approximation (Eq. (III.4)), compared to the Lifshitz-Slyozov
result for the locally conserved order parameter case (model $B$) [16][24].
Both distributions have been normalized and the $r$ axis is scaled such that
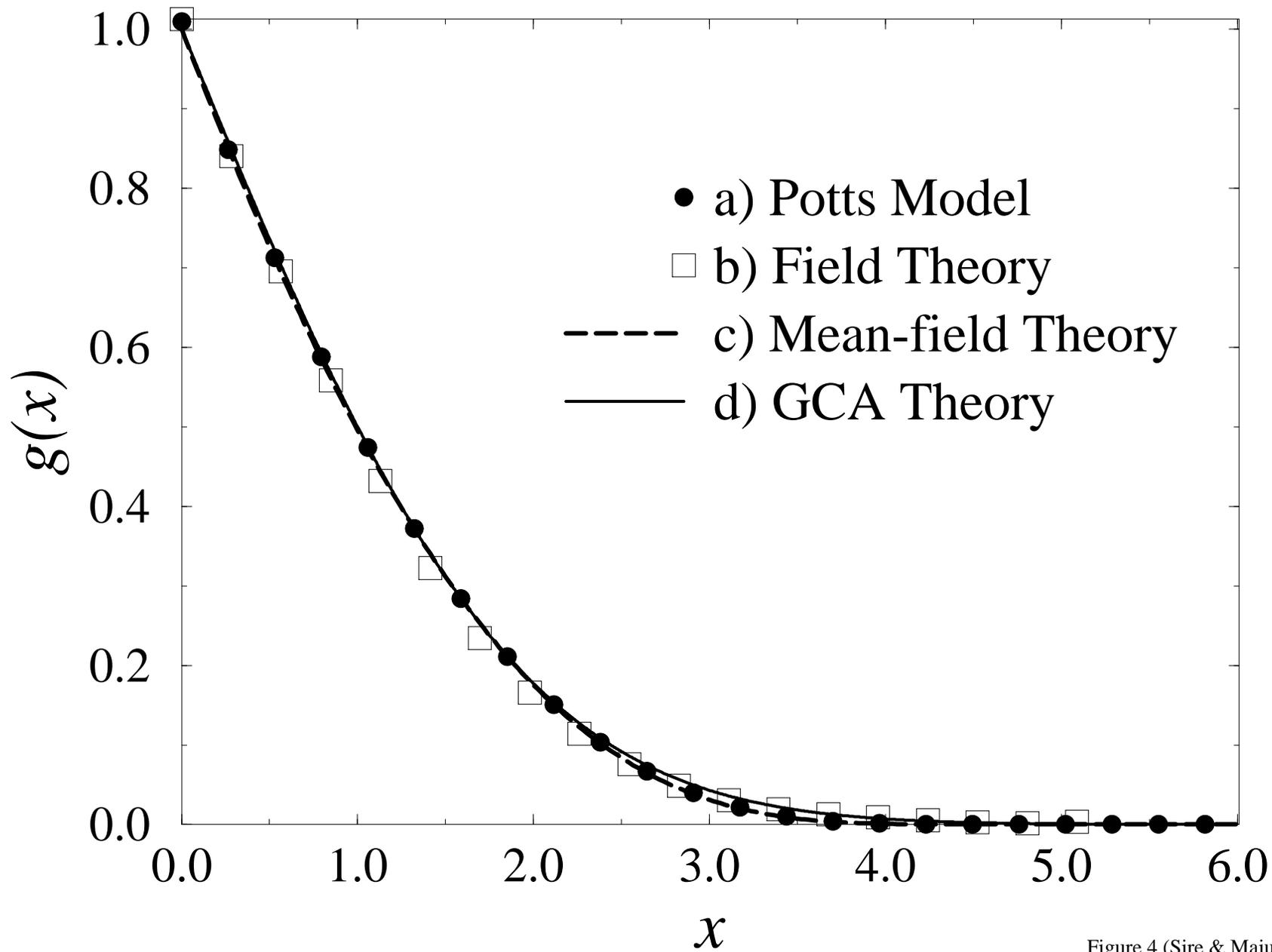
$\langle r\rangle =1$. \item{Fig. 4:} Comparison of the scaled equal-time
correlation functions generated by (a) numerical simulation of the Potts model
with $q=\infty$ (in fact $q$ equals the initial number of bubbles $N_{B}\sim
32000$), (b) numerical integration of Eq. (V.2) with $q=50$ (symbols have the
size of the typical error bars), (c) the mean-field theory of section III (Eq.
(III.7)), (d) the large $q$ GCA analytical result of section V.

\vskip 3cm \item{Tab. 1:} $\lambda$ for different $q$ as computed from the
Potts model simulation, for the globally conserved Ising model (see text), and
by solving the eigenvalue equation Eq. (V.15) numerically. $^1$ see also [9].
$^2$ see also [15].

\end